\documentclass{article}
\usepackage{mrs2005,epsfig,graphicx}
\begin{document} 
\title{Strong Line Metallicity Calibrators Applied to SDSS Galaxies}

\author{Sara L. Ellison}
\affil{University of Victoria, Dept. of Physics \& Astronomy, British
Columbia, Canada} 
 
\author{Lisa J. Kewley}
\affil{Institute for Astronomy, University of Hawaii, Hawaii, USA}
 
\begin{abstract} 
We calculate the oxygen abundances for $\sim$ 45,000 star-forming galaxies
selected from the SDSS using nine popular `strong-line' metallicity
diagnostics.  We find that individual galaxies can have metallicities 
which differ by
factors of up to $\sim$ 4, depending on the choice of abundance diagnostic.
The resulting mass-metallicity relations subsequently exhibit
significant offsets, with a range of 0.5 dex in metallicity at a given
mass.    This demonstrates that different metallicity diagnostics
should not be used interchangeably for determining oxygen abundances.
We also show that SDSS spectra are dominated by the central galactic
component which introduces a noticeable aperture effect.
\end{abstract} 
 
\section{Introduction} 

The metallicity of a galaxy is a powerful diagnostic which represents the
aggregate history of its star formation, mergers, gas infall and galactic
winds.  In other words, a single parameter encapsulates a complex
record of a galaxy's past.  For the most part, metallicities in nearby
galaxies are measured through their HII region gas phase oxygen abundances.
High S/N spectra of individual HII regions permit an explicit solution
for the electron temperature, density and, ultimately, metallicity.
Determination of metallicities `directly' in this way, typically 
requires detection of [OIII] $\lambda \lambda 4959, 5007$, H$\beta$ and 
[OIII] $\lambda$4363 (or some other auroral line).  As the 
metallicity of the HII region increases, far-IR lines become the
dominant coolant, so the optical auroral lines become increasingly weak.  
Very deep observations are therefore required to determine metallicities 
in regions
of even moderate metallicities in local galaxies (e.g. Bresolin et al. 2004).
Already at modest distances, the direct method of metallicity determination
from electron temperatures therefore becomes impractical due to the 
difficulty of detecting any of the weak auroral lines.  To circumvent 
this problem, a number of metallicity
diagnostics have been developed which make use of only the stronger
emission lines that can be readily observed out to even high redshifts
(see the contribution by M. Pettini in these proceedings).  These so-called
`strong-line' metallicity diagnostics have been calibrated in a variety
of ways, utilise different sets of emission lines and are convenient
for different redshift ranges.  In this proceedings contribution, I
describe work in which we have applied a range
of metallicity diagnostics to a large, uniform sample of galaxies
drawn from the SDSS and compare the metallicities derived in each case.
We find that the oxygen abundance derived for a given galaxy
can vary by up to a factor of $\sim 4$ depending on the choice of diagnostic.

\section{Galaxy Sample Selection} 

Our dataset was selected from the 261054-galaxy SDSS sample described in
Brinchmann et al. (2004) according to the following criteria:

\begin{enumerate}

\item Signal-to-noise (S/N) ratio of at least 3 in the strong
emission-lines H$\beta$, [OIII]~$\lambda 5007$,
H$\alpha$, [NII]~$\lambda 6584$, and [SII]~$\lambda \lambda 6717,31$.  This S/N
criterion is required for accurate classification of the galaxies into
star-formation or AGN-dominated classes e.g. Kewley et al (2001) and
Veilleux \& Osterbrock (1987).

\item Fibre covering fraction $> 20$\% of the total photometric g'-band
light.   Kewley, Jansen \& Geller (2005) found that a flux covering 
fraction above
this cut-off is
required for metallicities to begin to approximate global values.   Lower
covering fractions can produce significant discrepancies between
fixed-sized aperture and global metallicity estimates (although a 20 \%
covering fraction does not guarantee the absence of aperture effects,
as we discuss in section \ref{ap_sec}).

\item Stellar mass estimates must be available from the catalog of
Tremonti et al. (2004)
\end{enumerate}

The resulting sample contains 45086  galaxies,  spanning a large
range in stellar mass
($10^{8} - 10^{11.6}$ M$_{\odot}$) and metallicities $ 8.0<\log({\rm
O/H})+12<9.4$, as calculated by Tremonti et al. (2004).

\section{Metallicity Diagnostics}

%
%
\begin{figure}  
\centerline{\rotatebox{270}{\resizebox{10cm}{!}
{\includegraphics{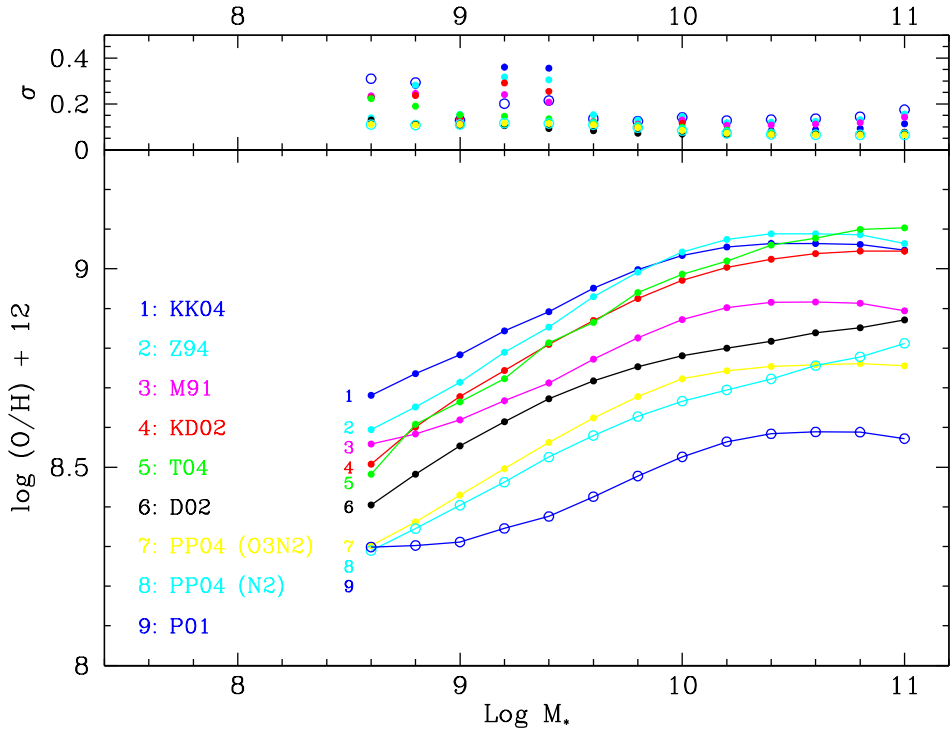}}}}
\caption{MZ relation for nine strong-line diagnostics binned by mass
in units of log M$_{\odot}$).
The top panel shows the RMS scatter for each calibration in each
mass bin.  }
\end{figure}

Strong-line metallicity diagnostics can be broadly separated into
three categories: theoretical, empirical and combination.  The
first category utilises theoretical models to calibrate emission
line flux ratios and includes the diagnostics of McGaugh (1991; M91),
Kewley \& Dopita (2002; KD02), Kobulnicky \& Kewley (2004; KK04),
Zaritsky, Kennicutt \& Huchra (1994; Z94) and Tremonti et al. (2004; T04).  
Empirical
methods rely on fitting a function to metallicities determined
via direct methods versus strong line flux ratios, such as Pilyugin
(2001; P01).  Finally, the combination methods again determine
a functional fit to data, but use a mix of direct
metallicity determinations, supplemented with values (often at the
high metallicity end) obtained via theoretical calibrations.  Examples
of combination calibrations include Denicolo et al. (2002; D02) and
the two calibrations of Pettini \& Pagel (2004; PP04).  
Each of these calibrations has associated
advantages and disadvantages, many of which are discussed in detail by
Kewley \& Ellison (2005).  For example, one of the disadvantages of a number
of diagnostics which use the so-called $R_{23}$ ratio of [OII]~$\lambda 3727$, 
[OIII]~$\lambda 4959, 5007$
and H$\beta$ is that there are two metallicity solutions for any
given flux ratio.  Moreover, the range in rest wavelengths of these
emission lines means that correction for internal galactic extinction
becomes very important for accurate flux ratios.
Although diagnostics which use the ratio of 
[NII]/H$\alpha$ (e.g. D02, PP04) do not suffer from these problems, 
they are sensitive
to the ionization parameter (the ratio of H ionizing photons to H atoms),
and effectively `saturate' at high metallicities.

We have applied the nine calibrations mentioned above to the sample
of star-forming galaxies described in \S2.  To investigate the 
importance of diagnostic choice, we focus on the impact manifested
in the mass-metallicity (MZ) relation recently studied by T04.
In Figure 1 we show the MZ relation for the nine strong-line diagnostics.
For clarity, we have binned the metallicities by mass and shown the
median values of each bin.  We see that
the MZ relations exhibit the same general trends for all diagnostics: a
rise in metallicity with mass up to a stellar mass log $M_{\star} \sim 10$
M$_{\odot}$ followed by a flattening at higher masses.
However, there is considerable difference in the normalization of the
MZ relation, with a range of 0.5 dex (a factor of 3) in metallicity for
a given mass between diagnostics.  Furthermore, in Figure 2 we show 
explicitly the comparison between two diagnostics, KK04 and P01, which 
shows that a given galaxy can have metallicities different by up to a
factor of $\sim 4$ depending on diagnostic choice.
This is a clear demonstration
that although different metallicity diagnostics may be convenient for
galaxies at different redshifts, consistency in metallicity
determination is imperative when combining samples.  If different
strong-line calibrations are used, this will, at best, introduce
a scatter in the MZ relation. At worst, a systematic offset could
be introduced, leading to erroneous comparisons between the samples.

%
%
\begin{figure}  
\centerline{\rotatebox{270}{\resizebox{10cm}{!}
{\includegraphics{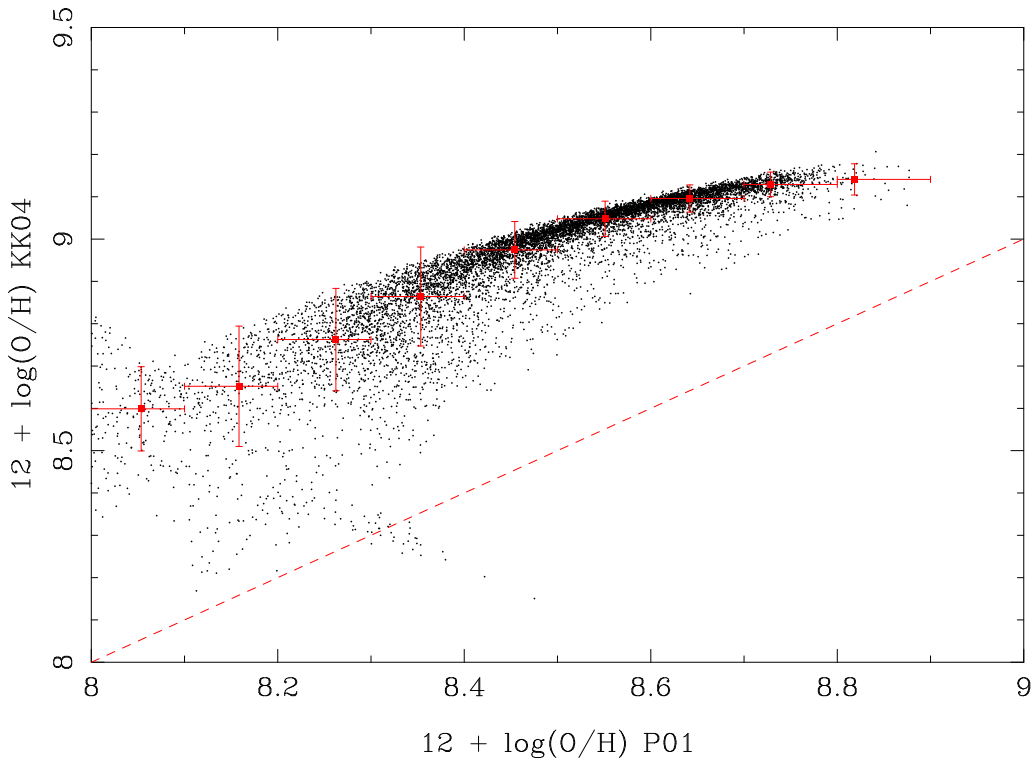}}}}
\caption{Oxygen abundances determined for SDSS galaxies using two different
strong-line diagnostics; Pilyugin (2001) and Kobulnicky \& Kewley (2004).
The red points with error bars are binned medians and the dashed line
shows a one-to-one relationship. }
\end{figure}

However, for some of the strong-line diagnostics, it may be possible
to correct for systematic calibration differences.  In a forth-coming
paper (Kewley \& Ellison 2005) we discuss the prospects
for such corrections and provide prescriptions for converting between
the diagnostics discussed in this proceedings contribution.

\section{Aperture Effects}\label{ap_sec}

%
%
\begin{figure}  
\centerline{\rotatebox{0}{\resizebox{10cm}{!}
{\includegraphics{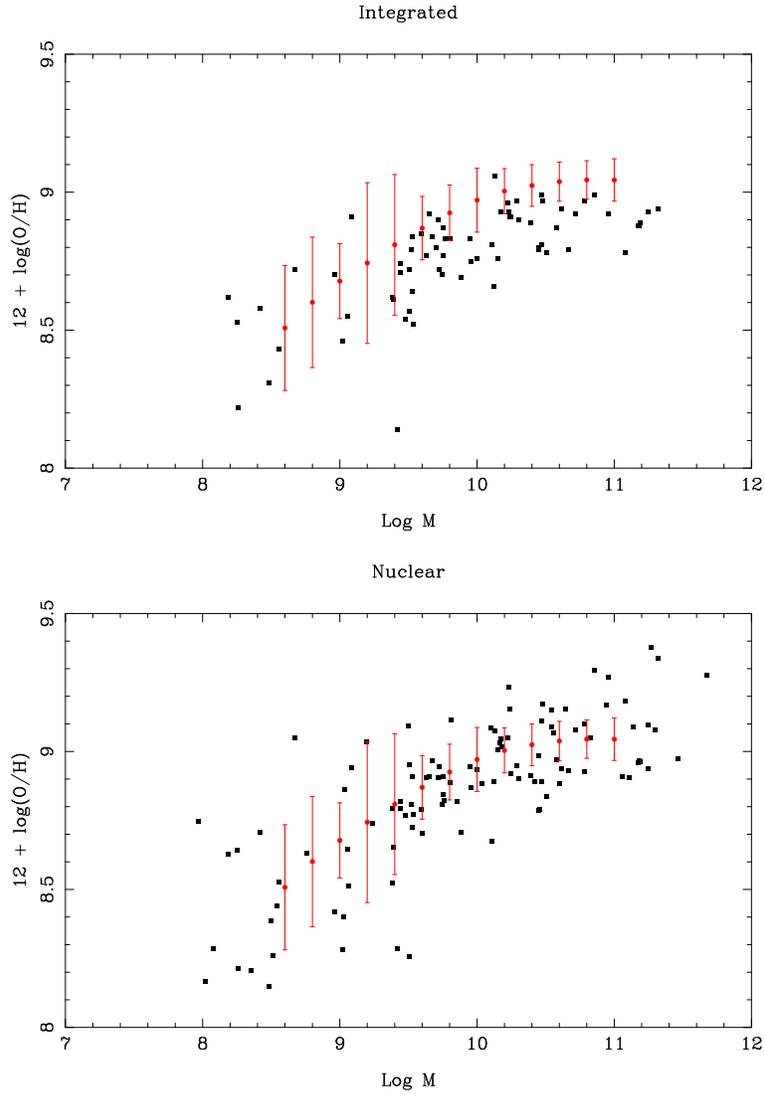}}}}
\caption{The SDSS MZ relation determined from the KD02 diagnostic, binned
by mass (red points with RMS error bars, both panels) compared with
the NFGS.  In the top panel, the SDSS data are compared with integrated
NFGS spectra (black points without error bars) which encompass the bulk 
of the galaxies' light.  In the lower panel, nuclear NFGS metallicities
were derived from slit spectra of just the central part of the galaxy. }
\end{figure}

We finish by briefly considering the effect of integrated fibre
spectroscopy on the derived `global' metallicity of a galaxy.  The
SDSS fibres are 3 arcsec in diameter, corresponding to a proper
length of 5.5 kpc at $z=0.1$ ($H_0$ = 70 km/s/Mpc, $\Omega_M = 0.3$,
$\Omega_{\Lambda} = 0.7$), the median redshift of our galaxy
sample.  What is the effect of only including the central component
of the galaxy's light in our spectroscopic determination of
metallicity?  Using the Nearby Field Galaxy Survey (NFGS) of
Jansen et al. (2000), Kewley et al. (2005) have suggested that
if the covering fraction is less than 20\%, aperture effects can
cause errors up to 40\% in the metallicities (and effects may persist
in high luminosity galaxies even for significantly larger covering 
fractions).  We investigate
the impact of aperture effects in SDSS data by returning to the
NFGS sample and determining its MZ relation.  We calculate NFGS
galaxy masses by combining 2MASS $J$ band magnitudes with optical
band passes, a technique which is comparable to determining the
mass via spectral fitting (Drory, Bender \& Hopp 2004).  In both panels
of Figure 3
we show again the MZ relation for the SDSS galaxies using the KD02
metallicity calibration.  In the upper panel, we compare the SDSS
data with KD02 metallicities for the \textit{integrated} metallicities
derived for the NFGS, calculated by combining slit spectra
taken at many positions across the face of the galaxy.  We see
that although the lower mass NFGS galaxies trace well the SDSS
MZ relation, the nearby galaxies appear to have a plateau at a
lower metallicity (by $\sim$ 0.15 dex).  In the lower panel of Figure 3,
we show the comparison of SDSS data with the \textit{nuclear}
NFGS metallicities, derived from a single long slit spectrum
taken across the centre of the galaxy.  Now we see a much better
agreement with the SDSS.  These results highlight how nuclear spectra
yield higher metallicities than integrated spectra, as expected
if strong abundance gradients are present.  Again, this is cause for caution
when comparing the SDSS galaxies with high redshift samples whose
spectra will capture more of the peripheral light than the
SDSS spectra which are weighted towards higher nuclear metallicities.
 
\section{Conclusions} 
 
This work has provided us with two important caveats to keep in
mind when dealing with galaxy metallicities.  First,
we have shown that considerable caution is required when dealing
with metallicities derived from HII region strong line diagnostics.  
Although it is impossible to advise on which diagnostic yields the
`right' answer, we have shown that one must be consistent with the 
choice of metallicity calibration.  Second,
we have demonstrated that SDSS spectra are subject to considerable
aperture effects.  Since a fixed sized aperture (slit or fibre) will
encompass different fractions of galaxy light as a function of redshift,
it is important to account for aperture effects before comparing SDSS
data with high redshift studies.


\vfill 
\end{document}